\documentclass[preprint,12pt]{elsarticle}

\usepackage{amssymb}
\usepackage{graphicx}
\usepackage{tabularx, multirow}
\usepackage{booktabs}
\usepackage{url}
\usepackage{amsmath}

\begin{document}

\begin{frontmatter}


\title{Adaptive Variance Thresholding: A Novel Approach to Improve Existing  Deep Transfer Vision Models and Advance Automatic Knee-Joint Osteoarthritis Classification}

\author[1]{Fabi Prezja}
\author[3,4]{Leevi Annala}
\author[1]{Sampsa Kiiskinen}
\author[1,5]{Suvi Lahtinen}
\author[1]{Timo Ojala}

\affiliation[1]{organization={University of Jyväskylä, Faculty of Information Technology},
            city={Jyväskylä},
            postcode={40014}, 
            country={Finland}}

\affiliation[3]{organization={University of Helsinki, Faculty of Science, Department of Computer Science},
            city={Helsinki},
            country={Finland}}

\affiliation[4]{organization={University of Helsinki, Faculty of Agriculture and Forestry, Department of Food and Nutrition},
            city={Helsinki},
            country={Finland}}

\affiliation[5]{organization={University of Jyväskylä, Faculty of Mathematics and Science, Department of Biological and Environmental Science},
            city={Jyväskylä},
            postcode={40014}, 
            country={Finland}}

\begin{abstract}
Knee-Joint Osteoarthritis (KOA) is a prevalent cause of global disability and is inherently complex to diagnose due to its subtle radiographic markers and individualized progression. One promising classification avenue involves applying deep learning methods; however, these techniques demand extensive, diversified datasets, which pose substantial challenges due to medical data collection restrictions. Existing practices typically resort to smaller datasets and transfer learning. However, this approach often inherits unnecessary pre-learned features that can clutter the classifier's vector space, potentially hampering performance. This study proposes a novel paradigm for improving post-training specialized classifiers by introducing adaptive variance thresholding (AVT) followed by Neural Architecture Search (NAS). This approach led to two key outcomes: an increase in the initial accuracy of the pre-trained KOA models and a 60-fold reduction in the NAS input vector space, thus facilitating faster inference speed and a more efficient hyperparameter search. We also applied this approach to an external model trained for KOA classification. Despite its initial performance, the application of our methodology improved its average accuracy, making it one of the top three KOA classification models.
\end{abstract}


\end{frontmatter}


\newcommand{\qvec}[1]{\textbf{\textit{#1}}}

\section{Introduction}
Over the last ten years, a significant increase has been observed in the incorporation of artificial intelligence into the medical field\cite{wang2019deep,beam2018big}, spurred by the robust expansion in deep machine learning methodologies\cite{lecun2015deep}. Medicine has been recognized as a key domain for deploying these sophisticated technologies, with deep learning primarily focused on data analysis and clinical decision support. These proficient systems, capable in exploring medical data to identify patterns and relationships, cover many applications. They've displayed considerable advancement in predicting patient outcomes \cite{kather2019predicting,courtiol2019deep,diamant2019deep}, along with enriching diagnostics and disease classification \cite{esteva2017dermatologist,han2017breast,bakator2018deep,prezja2023improved,prezja2023improving}. Apart from data analysis and classification, deep learning has been successful in data segmentation \cite{isensee2021nnu,liu2021review}, and has also shown progress in the generation \cite{chuquicusma2018fool,calimeri2017biomedical,frid2018gan,thambawita2021deepfake,annala2020generating} and anonymization of medical data\cite{shin2018medical,yoon2020anonymization,torfi2022differentially,kasthurirathne2021generative,prezja2022deepfake,prezja2022synthetic}. Nonetheless, applying these technological developments to Osteoarthritis (OA) brings its unique set of challenges.

OA, predominantly knee joint osteoarthritis\cite{yeoh2021emergence,saarakkala2010depth, laasanen2003biomechanical} (KOA), is a leading cause of disability worldwide\cite{hermans2012productivity}, with estimated expenses amounting to as much as 2.5\% of the gross national product in western countries.\cite{hermans2012productivity}. Early detection is hampered by subtle radiographic indicators and variability in disease progression\cite{HUNTER20191745,yeoh2021emergence}. Employing deep learning in KOA classification \cite{tiulpin2019multimodal,tiulpin2018automatic,tiulpin2020automatic} heavily relies on diverse and extensive data sets. Yet, obtaining such data sets is challenging, hindered by patient privacy concerns\cite{centers2003hipaa,voigt2017eu}, limitations in data collection, and the intrinsic progression of OA. Traditionally, researchers have circumvented these limitations through deep transfer learning, where pre-trained neural networks are repurposed for a specific task using smaller data sets. The peak accuracy achieved for multi-KL classification using radiographic Kellgren and Lawrence grading of OA stands at 74.81\% \cite{zhang2020attention,yeoh2021emergence}. This process typically involves end-to-end training of transfer architectures, which include a feature learning section and a classification section. In this setup, learned features are usually fed into a classifier for predictions. However, not all repurposed features may be highly relevant to the task, potentially cluttering the classifier with unnecessary or barely varying (near-static) features, while manual classifier design could increase complexity and raise the risk of under-fitting. However, feature selection\cite{khalid2014survey} and Neural Architecture Search (NAS)\cite{elsken2019neural,JMLR:v24:20-1355} can partly mitigate these issues. Feature selection refining classifier input and NAS automating the architecture design process, identifying optimal structures often missed in manual design, thus increasing potential model efficiency and accuracy. However, these techniques have yet to be utilized with transfer learning-based classification for KOA classification.

In this study, we propose a method to amplify the performance of existing KOA classification models using an adaptive variance threshold feature filtering technique and neural architecture search. This approach refined the classifier's feature space, boosted computational efficiency, and advanced the initial accuracy of the evaluated models. Significantly, when we applied our technique to an external model, specifically from Chen et al\cite{chen2019fully}, its average accuracy improved to 71.14\%.
This boost propeled the external model to the top three KOA classification models\cite{yeoh2021emergence,zhang2020attention}.

\section{Methods}

\begin{figure}[!ht]
\centering
\includegraphics[width=\textwidth]{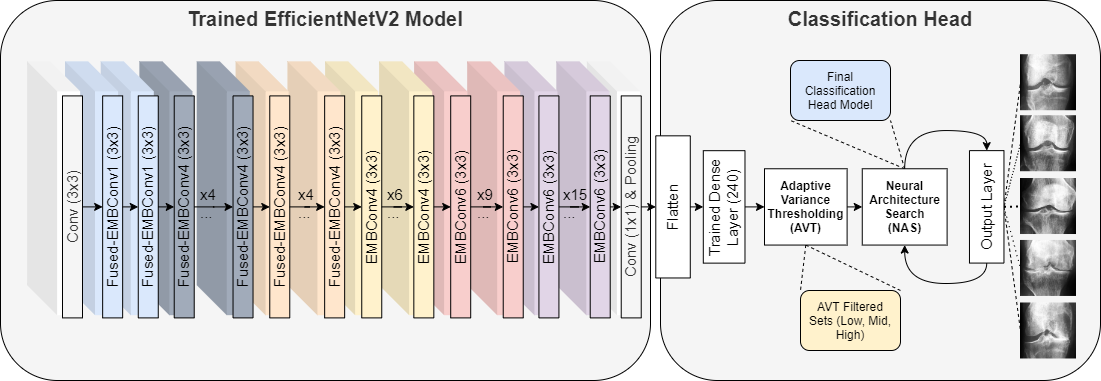}
\caption{The study's methodological pipeline, as applied to the EfficientNetV2 Base Model, showcases the intervention in the classification head.}
\label{flow}
\end{figure}

Our study's methodology is structured into three clear phases. First, we focus on data collection and essential pre-processing. In the subsequent phase, we explore the intricacies of the deep transfer learning approach. The concluding phase offers a comprehensive overview of the adaptive variance thresholding method and neural architecture search, detailing the metrics for evaluation. Figure \ref{flow} illustrates the core approach we used with the pre-trained EfficientNet model for KOA classification.

\subsection{Data Collection} 
\begin{figure}[!ht]
\centering
\includegraphics[width=\textwidth]{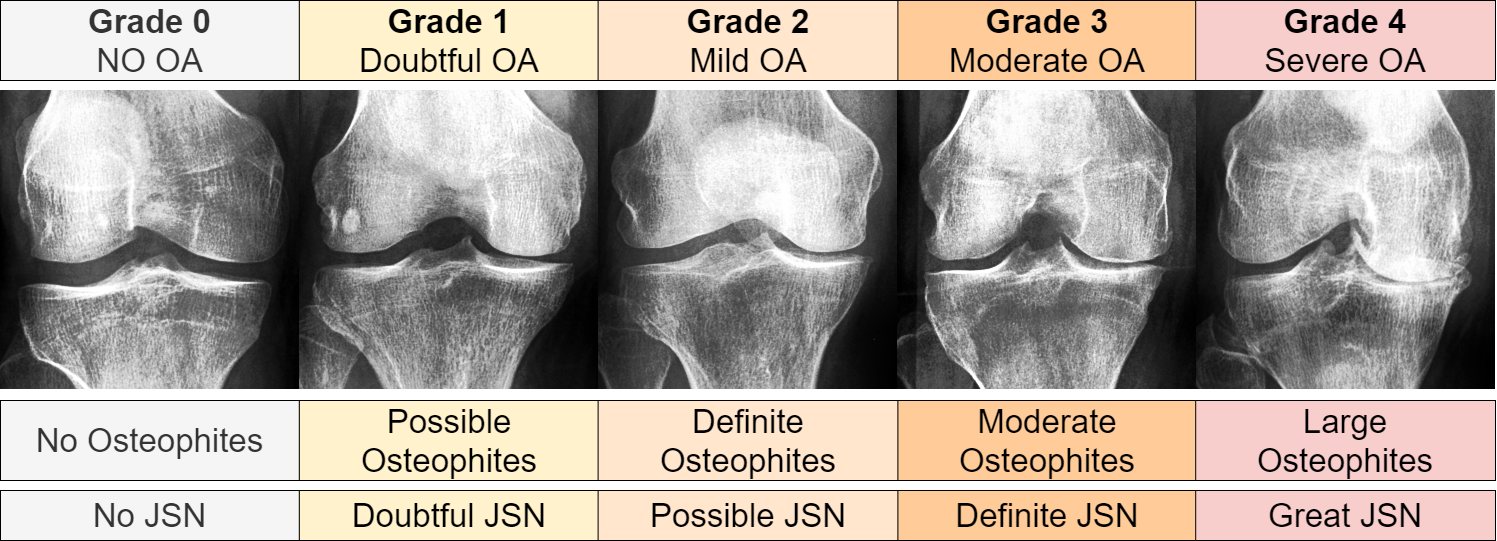}
\caption{Sample images display different KL grades, from 0 (indicating no OA signs) to 4 (representing severe OA). As we move from left to right, the severity of OA intensifies. Within each image, two red markers are present: a circle pinpointing osteophyte areas and an arrow highlighting joint space narrowing (JSN).}
\label{grades}
\end{figure}

Our study utilized knee joint X-ray images derived from the 2019 Chen, et al. study\cite{chen2019fully}, obtained initially from the Osteoarthritis Initiative (OAI). The OAI is a multi-center study focused on biomarkers for knee osteoarthritis that involved 4796 participants aged between 45 and 79. We used the pre-processed primary cohort data\cite{chen2018knee} from the Chen 2019 study, which had undergone automatic knee joint detection, bounding, and standardization of zoom to 0.14mm/pixel. This process yielded 8260 images (224 x 224 pixels) extracted from 4130 X-rays, each encompassing both knee joints. The images were classified using the Kellgren and Lawrence (KL) system\cite{kellgren1957radiological}, as illustrated in Figure \ref{grades}. The distribution of KL grades comprised 3253 images for Grade 0, 1495 for Grade 1, 2175 for Grade 2, 1086 for Grade 3, and 251 for Grade 4.

\subsection{Image Pre-Processing}

We adjusted each right knee joint image to replicate the orientation of a left knee. We subsequently located and inverted any negative channel images, leading to 189 instances. The contrast of the image histograms was then equalized using equation \ref{eq:contr_norm}. In this equation, we took a given grayscale image $\qvec{I}$ of $m\times n $ dimensions and used the cumulative distribution function $cdf$ and pixel value $v$ to derive an equalized value $h(v)$ within the range $[0,255]$. Here, $cdf_{min}$ represents a non-zero minimum value of the image's cumulative distribution, while $m\times n$ refers to the total number of pixels.

\begin{equation}\label{eq:contr_norm}
h(v)=255\frac{cdf(v)-cdf_{min}}{(mn)-cdf_{min}}
\end{equation}

\subsection{Chen 2019 External Validation}

In order to externally validate our approach, we mirrored the exact models  and data utilized in the Chen et al. 2019 study\cite{chen2019fully}. We chose the study's best-performing VGG19 - Ordinal model (publicly available\cite{chen2018knee}). We extracted the feature vectors from this model from the pre-output dense layer, effectively replicating the conditions under which the original model achieved its top performance. This step was crucial to ensure a fair comparison and assessment of the effectiveness of our approach when applied to an external model.

\subsection{Convolutional Neural Networks}
Convolutional Neural Networks (CNNs) \cite{lecun1995convolutional}, a cornerstone in the renaissance of deep learning \cite{lecun2015deep}, have often found applications in the realm of computer vision. The strength of CNNs lies in their ability to execute convolution operations between an input and a filter-kernel, thereby highlighting distinctive features captured in a response known as a feature map. As these filters slide across inputs, they create layers of complex feature maps, each representing more abstract concepts from the preceding maps. Mathematically, given an image $\qvec{I}$ with dimensions $u\times v$ and a filter-kernel $\qvec{H}$ of $s\times t$ dimensions, the generation of a feature map $\qvec{G}$ by convolving $\qvec{I}$ and $\qvec{H}$ across axes $u, v$ is expressed as:

\begin{equation}\label{eq:cnn}
\qvec{G}(u,v)=\sum_{s}\sum_{t}\qvec{I}(u,v)\qvec{H}(u-s,v-t)
\end{equation}

Usually, the resulting values of the feature map are filtered through an activation function, which acts to re-map these values using a pre-defined function. One common example is the Rectified Linear Unit activation function\cite{nair2010rectified} (ReLu), which sets negative values to zero, thereby offering computational efficiency by eliminating less essential information. For any feature map value $z$, the ReLu activation function is defined as:

\begin{equation}\label{eq:relu}
f(z)= \max (0,z)
\end{equation}

Further, a max pooling operation is often employed to down-sample the convolution result. Consequently, cascades of max pooling and convolution yield a gradually decreasing length of features. For an image $\qvec{I}$ with dimensions $u\times v$, the max pooled value ${g(u_I)}$ for dimension $u$ can be defined as:

\begin{equation}\label{eq:maxp}
g(u_I)= \lfloor\frac{{u_I}-r}{h}\rfloor+1
\end{equation}

Here, $u_I$ is dimension $u$ from image $\qvec{I}$, $r$ denotes the pooling window size, and $h$ signifies the stride value. This streamlining process contributes to an efficient and effective model for identifying the necessary classification task features.

\subsection{Convolutional Neural Network Architecture}

A notable deep learning model, EfficientNet\cite{tan2019efficientnet} , adopts the strategy of compound scaling, which systematically adjusts depth (number of layers), width (size of the layers), and resolution (size of the input image). This scaling process is mathematically encapsulated as follows:

\begin{equation}\label{eq:enet}
d = \alpha^{\phi} d_0, \quad w = \beta^{\phi} w_0, \quad r = \gamma^{\phi} r_0
\end{equation}

Here, $\alpha, \beta, \gamma$ denote constants, $\phi$ stands for a user-selected coefficient, and $d_0, w_0, r_0$ represent the depth, width, and resolution of the original model, respectively.

A distinctive feature of EfficientNet is the MBConv block, which chains together transformations in the order of a $1\times1$ convolution, a depth-wise convolution, a Squeeze-and-Excitation (SE) operation \cite{hu2018squeeze}, and a subsequent $1\times1$ convolution:

\begin{equation}\label{eq:mbconv_se}
T_{MB}(\qvec{I}) = \qvec{K}_2 \ast SE(\qvec{D} \ast (\qvec{K}_1 \ast \qvec{I}))
\end{equation}

In this equation, $\qvec{K}_1$ and $\qvec{K}_2$ are $1\times1$ convolutional filters, $\qvec{D}$ stands for the depth-wise convolutional filter, and $SE(\cdot)$ signifies the Squeeze-and-Excitation operation. The EfficientNetV2\cite{tan2021efficientnetv2} model extends its predecessor by incorporating a Fused-MBConv block that fuses the initial $1\times1$ and depth-wise convolutions into a single $3\times3$ convolution, followed by an SE operation and a concluding $1\times1$ convolution:

\begin{equation}\label{eq:fmbconv}
T_{FMB}(\qvec{I}) = \qvec{K}{2} \ast (SE(\qvec{K}{f} \ast \qvec{I}))
\end{equation}

\begin{figure} [!ht]
\centering
\includegraphics[width=\textwidth]{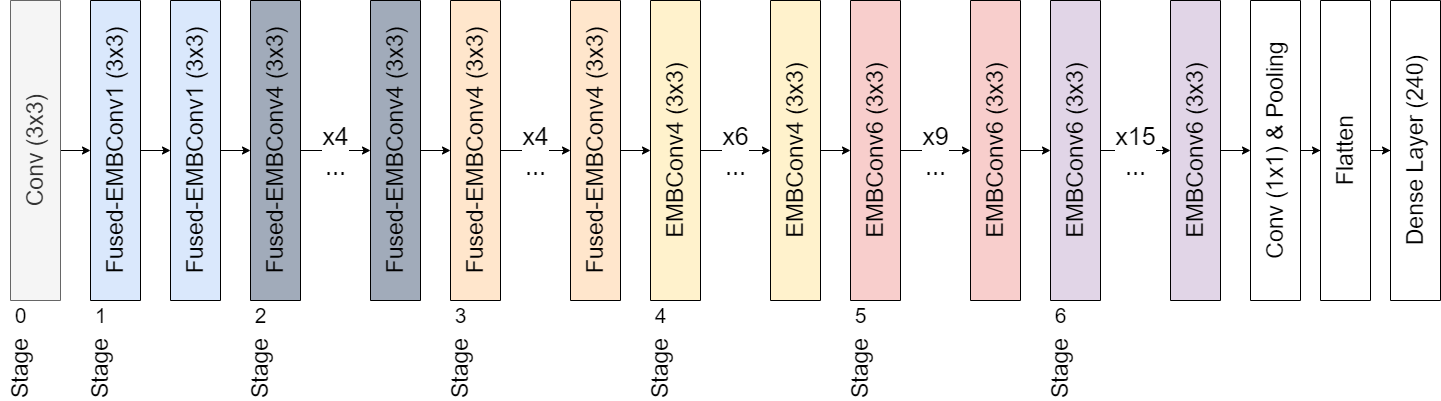}
\caption{The base architecture of EfficientNetV2, complemented by the changes we implemented after the pooling stage.}
\label{fig:effinet}
\end{figure}

Here, $\qvec{K}_{f}$ is the $3\times3$ convolutional filter that merges the initial $1\times1$ and depth-wise convolutions, and $\qvec{K}_{2}$ is the final $1\times1$ convolutional filter. An activation function follows each convolution and may include a skip connection. The EfficientNetV2 base model (B0) and the modifications implemented in our study are illustrated in Figure \ref{fig:effinet}.

We utilized the EfficientNetV2-M architecture for our model and further updated it by introducing flattening and a dense layer consisting of 240 neurons. The model was trained over 25 epochs using the Adam optimizer, with the dataset being partitioned into Training (75\%), Validation (15\%), and Testing (15\%) sets (patient-wise splits). The minimum validation loss governed the early stopping criterion. The CNN training and online affine augmentations ('advanced augmentation preset') were executed using the open-source Deep Fast Vision library\cite{fabprezja_2023}.

\subsection{Adaptive Variance Thresholding}

Variance Thresholding is a popular feature selection method\cite{scikit-learn_variance_threshold_code,kuhn2013applied} used to enhance machine learning model performance by removing excess features. It calculates the variance of each feature and drops those whose variance is below a specific user-defined threshold. However, the conventional method requires manually setting the threshold, which is not adaptable to different datasets and does not generalize across datasets or feature vectors of different architectures trained even on the same dataset. To overcome this, we propose Adaptive Variance Thresholding (AVT), which automatically sets the threshold from a user-defined percentile of the calculated feature variances. This approach makes the method flexible, as it adapts to the unique characteristics of each dataset without requiring manual changes. The following mathematical steps describe the process of Adaptive Variance Thresholding (AVT):

\begin{enumerate}

\item Compute the variance of each feature in the feature set, $\qvec{F}$. We denote this variance as $\qvec{v}$, where each element $v_i$ represents the variance of the $i$-th feature:
\begin{equation}
\mathbf{v} = \text{Var}(\qvec{F}_{i}), \quad \forall i \in {1, 2, ..., w}
\end{equation}

Here, $\qvec{F}_{i}$ denotes the $i$-th feature in the feature matrix $\qvec{F}$ with dimensions $w \times z$ (where $w$ is the number of features and $z$ is the number of samples). $\text{Var}(\qvec{F}_{i})$ calculates the variance of the $i$-th feature, with $\mathbf{v}$ storing these variances.

\item Set the threshold $j$ to be the $p$-th percentile of these computed variances:
\begin{equation}
j = \text{percentile}(\mathbf{v}, p)
\end{equation}

The function $\text{percentile}(\mathbf{v}, p)$ computes the $p$-th percentile of the variances in $\mathbf{v}$, where $p$ is a user-defined parameter.

\item Form a new feature matrix $\qvec{F'}$, consisting of only those features $i$ whose variance exceeds the threshold $j$:

\begin{equation}
\qvec{F}_i \in \qvec{F'} \text{ if } v_i \geq j
\end{equation}

\end{enumerate}

In the final step, the feature $\qvec{F}_{i}$ is included in the new feature matrix $\qvec{F'}$ if its corresponding variance $v_{i}$ is greater than or equal to the threshold $j$. Otherwise, the feature is removed. The power of the adaptive approach is in its capacity to derive the threshold directly from the data, allowing it to adjust to the distinct characteristics of each dataset. In our study, we examined three threshold levels, referred to as Low, Mid, and High, each representing different percentiles - 1.5\%, 50\%, and 98.5\%, respectively, given the computational demands associated with this approach; conducting an exhaustive search was prohibitively computationally demanding. In essence, we preserved the top 98.5\% varying features in the Low setting, the top 50\% in the Mid setting, and only the top 1.5\% in the High setting. These settings explored various scenarios from low to significant dimensionality reduction. The Python class implementing this method can be found in the provided materials. It constitutes a modification of the Scikit-Learn class\cite{scikit-learn}, using NumPy\cite{harris2020array} for computations.

\subsection{Neural Architecture Search}

Neural Architecture Search (NAS) \cite{elsken2019neural}, a method in machine learning, offers an automated way to design neural network architectures, diminishing the need for exhaustive manual tuning. By systematically exploring a range of potential architectures, NAS pinpoints the ones that yield the best performance for a specific task. The process involves defining a search space of potential architectures and using a controller model to generate, assess, and train these candidate architectures. In our experiment, we applied the AutoKeras\cite{JMLR:v24:20-1355} Structured Classifier search approach to identify suitable architectures for the Adaptive Variance Thresholding (AVT) sets. Notably, the evaluation of NAS was performed exclusively using validation data. We permitted a maximum of 55 consecutive trials with validation accuracy guiding the early stopping criterion. Each trial was given a training space of up to 25 epochs. All identified architectures are detailed in the appendix, and trained versions of these architectures are available under data availability.

\subsection{Classification Metrics}

Accuracy, a common measure of model performance, is the proportion of correct predictions (true positives and true negatives) to the total number of observations. It's mathematically defined as:

\begin{equation}
\text{Accuracy} = \frac{TP + TN}{TP + TN + FP + FN}
\end{equation}

Precision, also referred to as the positive predictive value, denotes the fraction of correct positive predictions in relation to all positive predictions made. The calculation for precision is as follows:

\begin{equation}
\text{Precision} = \frac{TP}{TP + FP}
\end{equation}

Recall, alternatively known as sensitivity or true positive rate, represents the fraction of actual positive instances that the model correctly identified. The formula for recall is:

\begin{equation}
\text{Recall} = \frac{TP}{TP + FN}
\end{equation}

The F1 score aims to balance precision and recall by taking their harmonic mean. It's calculated using the following formula:

\begin{equation}
\text{F1 Score} = 2 \cdot \frac{\text{Precision} \cdot \text{Recall}}{\text{Precision} + \text{Recall}}
\end{equation}

In these expressions, TP stands for True Positives, TN represents True Negatives, FP is for False Positives, and FN refers to False Negatives.

\section{Results}

\subsection{External Chen 2019 Model}

\begin{figure} [!ht]
\centering
\includegraphics[width=\textwidth]{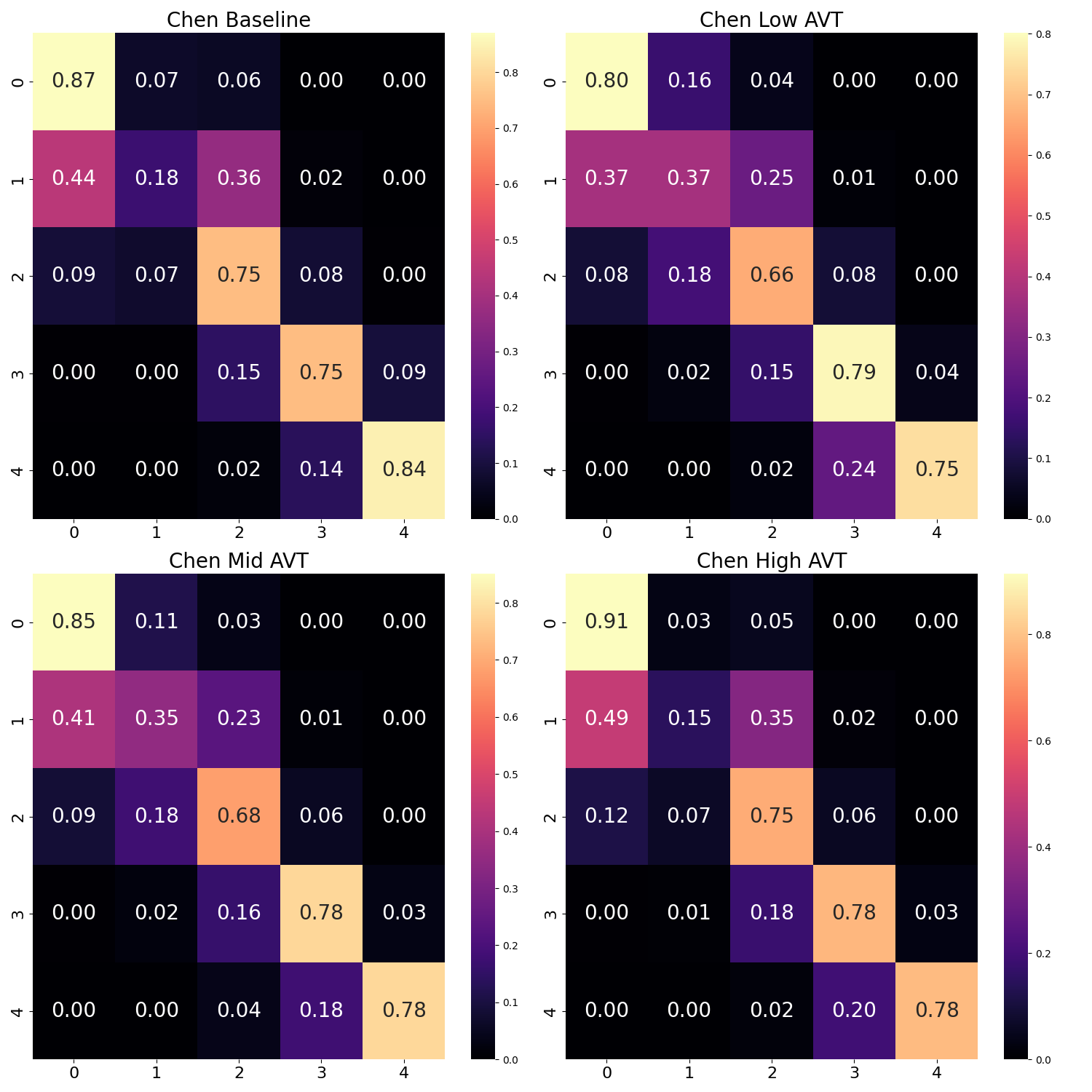}
\caption{The confusion matrices for the external Chen model using Chen test data. The baseline indicates no intervention, whereas Low, Mid, and High represent AVT sets. The appendix tables \ref{chenlowavtnas},\ref{chenmidavtnas},\ref{chenhighavtnas} detail the NAS architecture discovered for each set. X-axis shows predicted labels; Y-axis shows true labels.}
\label{fig:chenconf}
\end{figure}

This study utilized adaptive variance thresholding (AVT) and neural architecture search on pre-trained vision neural networks. Table \ref{chenrCFtab} compares the performance across AVT thresholds on the best pre-trained Chen 2019 model (baseline), highlighting the impact of these techniques. We observed that the model's NAS input dimensionality significantly decreased from 4096 in the 'Baseline' condition to 62 in the 'High AVT' condition. Accuracy generally improved, peaking at 71.14\% in the 'High AVT' scenario. Precision increased to 70.98\% in the 'Mid AVT' condition and slightly declined to 69.76\% in the 'High AVT' setting. Recall remained relatively steady across all conditions with a minor increase to 68.95\% in the 'Mid AVT' scenario. The F1-score, crucial in handling uneven KL distributions, increased until 'Mid AVT', then slightly decreased in the 'High AVT' condition. Figure \ref{fig:chenconf} further details the confusion matrixes for each condition.

\begin{table}[!ht]
\caption{Classification report for the Chen model using Chen test data, with and without AVT intervention (baseline). 'Dimensionality' denotes the vector space size pre-NAS.}
\label{chenrCFtab}
\small
\begin{tabularx}{\textwidth}{lXXXX}
\toprule
\textbf{} & \textbf{Baseline} & \textbf{Low AVT} & \textbf{Mid AVT} & \textbf{High AVT} \\
\midrule
Dimensionality & 4096 & 4034 & 2048 & 62 \\
Accuracy & 69.63\% & 68.30\% & 70.41\% & \textbf{71.14\%} \\
Precision & 64.79\% & 68.71\% & \textbf{70.98\%} & 69.76\% \\
Recall & 67.74\% & 67.32\% & \textbf{68.95\%} & 67.53\% \\
F1-score & 64.96\% & 67.96\% & \textbf{69.86\%} & 66.88\% \\
\bottomrule
\end{tabularx}
\end{table}

In the provided confusion matrices and accuracy results, the impact of varying Adaptive Variance Thresholding (AVT) levels, namely Baseline, Low AVT, Mid AVT, and High AVT, on the performance of a neural network model is presented. A general improvement in accuracy was observed as the AVT level increased, rising from approximately 69.63\% at Baseline to 71.14\% at High AVT. A detailed look into the normalized confusion matrices revealed a consistent rise in correct predictions, notably for the first and third classes, across increasing AVT levels. An exception was observed for KL2, where correct predictions declined as AVT increased. On the other hand, the fourth and fifth classes exhibited stability in model performance with varying AVT levels after an initial improvement from Baseline to Low AVT.

Different trends surfaced when we focused on the recall values for individual classes (as presented in Table \ref{chenrecall}). The model's proficiency in identifying instances of KL 0 consistently strengthened with increasing AVT, leading to the highest recall across all classes and conditions at High AVT (91.39\%). Conversely, KL 1 experienced a peak in recall at Low AVT (36.82\%), followed by a decrease at High AVT (14.86\%), lower than its Baseline value. KL 2 also benefited from higher AVT levels, as indicated by a higher recall at High AVT (75.39\%), despite slight decreases at Low and Mid AVT. The KL 3 class maintained a relatively stable recall across all AVT levels, peaking slightly at Low AVT (78.92\%). Finally, the recall for KL 4 initially dropped from Baseline to Low AVT but then held relatively steady for Mid and High AVT. These recall trends underline the complex interplay between AVT levels and KL grade-specific performance.

\begin{table}[!ht]
\caption{Recall report for the external Chen model across classes, derived from the confusion matrix diagonals.}
\label{chenrecall}
\small
\begin{tabularx}{\textwidth}{lXXXX}
\toprule
\textbf{KL Label} & \textbf{Baseline} & \textbf{Low AVT} & \textbf{Mid AVT} & \textbf{High AVT} \\
\midrule
0 & 87.01\% & 80.13\% & 85.13\% & \textbf{91.39\%} \\
1 & 17.57\% &\textbf{ 36.82\%} & 35.14\% & 14.86\% \\
2 & 74.94\% & 66.22\% & 68.01\% & \textbf{75.39\%} \\
3 & 74.89\% & \textbf{78.92\%} & 78.03\% & 77.58\% \\
4 & \textbf{84.31\%} & 74.51\% & 78.43\% & 78.43\% \\
\bottomrule
\end{tabularx}
\end{table}

\subsection{EfficientNetV2M Internal Model}

The following section will focus on a different neural network model - the EfficientNetV2M. When looking at Table \ref{efinetCFR}, it was clear that the dimensionality of the NAS input reduced dramaticaly as we moved from the Baseline to the High AVT level. Regarding model performance, Accuracy, and Precision peaked at the Mid AVT level, with Accuracy reaching 65.48\% and Precision achieving 67.18\%. However, Precision took an upward leap at High AVT, reaching 79.72\%, the highest among all levels. In contrast, Recall and F1-score exhibited a slight downward trend after the Baseline, with the lowest values observed at the High AVT level - 58.99\% and 58.35\%, respectively. Based on Accuracy and Precision, the model performance seemed optimal at Mid AVT rather than High AVT, suggesting that extreme dimensionality reduction might not necessarily lead to better overall performance for this set-up.

\begin{table}[!ht]
\caption{Classification report for the EfficientNet model using the test data, with and without AVT intervention (baseline). 'Dimensionality' denotes the vector space size pre-NAS.}
\label{efinetCFR}
\small
\begin{tabularx}{\textwidth}{lXXXX}
\toprule
\textbf{} & \textbf{Baseline} & \textbf{Low AVT} & \textbf{Mid AVT} & \textbf{High AVT} \\
\midrule
Dimensionality & 240 & 236 & 120 & 4 \\
Accuracy & 63.23\% & 64.27\% & 65.48\% & 65.08\% \\
Precision & 66.25\% & 65.15\% & 67.18\% & 79.72\% \\
Recall & 66.24\% & 63.57\% & 64.12\% & 58.99\% \\
F1-score & 65.36\% & 63.25\% & 64.15\% & 58.35\% \\
\bottomrule
\end{tabularx}
\end{table}

\begin{figure} [!ht]
\centering
\includegraphics[width=\textwidth]{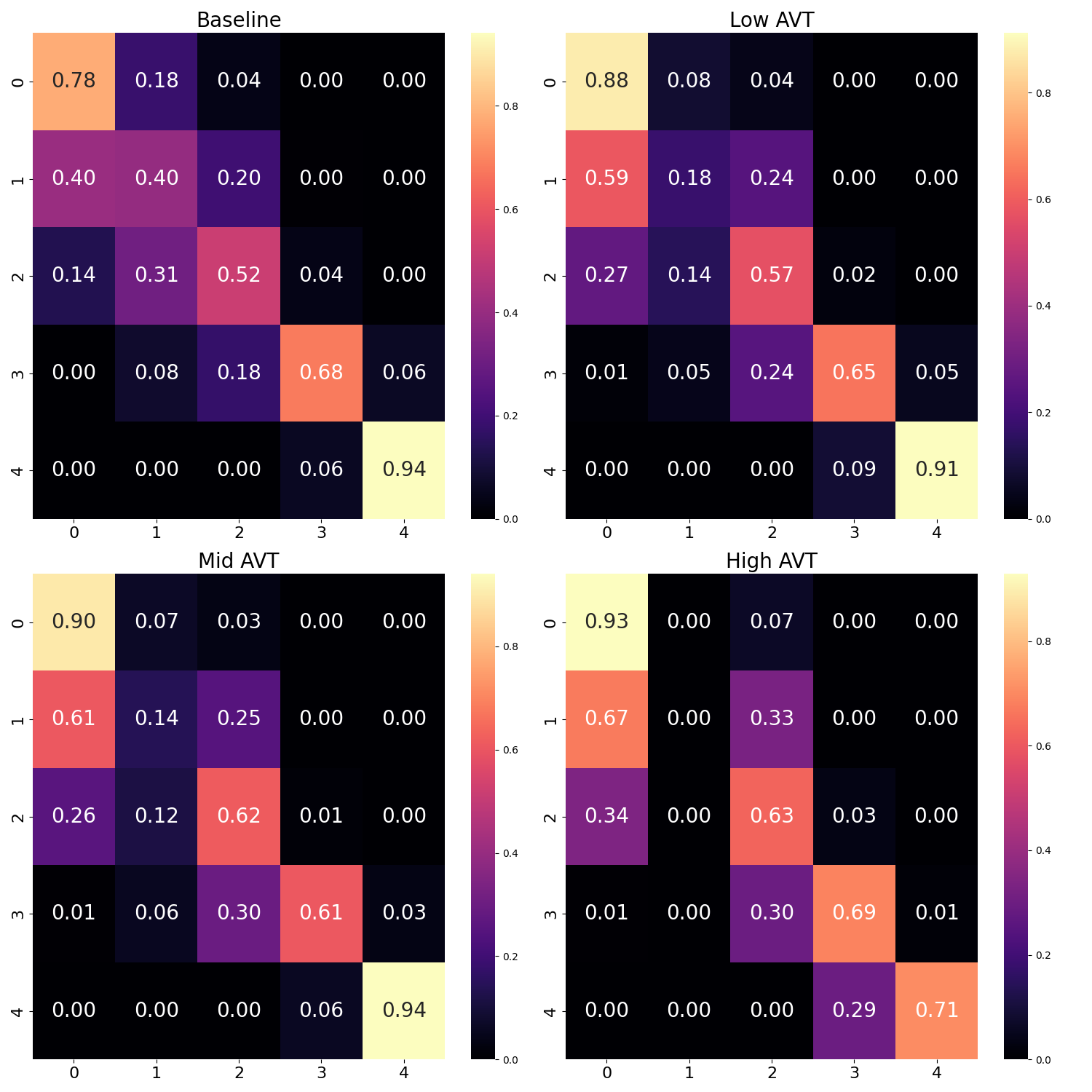}
\caption{The confusion matrices for the EfficientNet model using the test data. The baseline indicates no intervention, whereas Low, Mid, and High represent AVT sets. The appendix tables \ref{efilowavt},\ref{efimidavt},\ref{efihighavt} detail the NAS architecture discovered for each set. X-axis shows predicted labels; Y-axis shows true labels.}
\label{fig:ourconf}
\end{figure}

When analyzing the confusion matrices (Figure \ref{fig:ourconf}), we observe an overall upward trend in the diagonal values, indicating an increase in correct predictions on a higher variance threshold (AVT). Notably, KL 0 and KL 2 experienced consistent improvement in accuracy. The performance, however, varied for other classes. While KL 1's score declined post-Baseline and droped to zero at High AVT, KL 3 showed a slight inconsistency, and KL 4 displayed a fluctuating trend, peaking at Baseline and Mid AVT but falling at High AVT. The off-diagonal elements reveal changes in KL confusion across different AVT levels. Mainly, confusion between KL 0 and KL 1 decreased with an increasing AVT, while the confusion between KL 1 and KL 2 increased. The recall table \ref{ourrecaltab} further supplements these findings.

\begin{table}[!ht]
\caption{Recall report for the EfficientNet model across classes, derived from the confusion matrix diagonals.}
\label{ourrecaltab}
\small
\begin{tabularx}{\textwidth}{lXXXX}
\toprule
\textbf{KL Label} & \textbf{Baseline} & \textbf{Low AVT} & \textbf{Mid AVT} & \textbf{High AVT} \\
\midrule
0 & 77.66\% & 87.68\% & 89.98\% & 92.90\% \\
1 & 39.51\% & 17.56\% & 14.15\% & 0.00\% \\
2 & 51.73\% & 56.65\% & 61.56\% & 62.72\% \\
3 & 68.18\% & 64.77\% & 60.80\% & 68.75\% \\
4 & 94.12\% & 91.18\% & 94.12\% & 70.59\% \\
\bottomrule
\end{tabularx}
\end{table}

\clearpage

\section{Discussion}

This study evaluated a novel method for enhancing the performance of pre-existing models used for diagnosing Knee-Joint Osteoarthritis (KOA) using a combination of adaptive variance thresholding (AVT) and neural architecture search (NAS). This effort led to a notable increase in diagnostic accuracy, a substantial reduction in the NAS input vector space, and an overall improvement in model efficiency. More specifically, our approach improved model accuracy on the Chen KOA model, from 69.63\% at the baseline to 71.14\% at High AVT. This approach rendered the new solution within the top three radiographic KOA classification solutions \%\cite{yeoh2021emergence,zhang2020attention}. We observed that this enhancement was exceptionally beneficial for KL 0 and KL 2, showing a consistent increase in recall as the AVT level increased. Our internal model, EfficientNetV2M, also benefited from the new method. Although the overall model performance peaked at Mid AVT, we observed a remarkable jump in Precision at High AVT, suggesting that extreme dimensionality reduction could still lead to specific performance improvements. However, it is worth noting that while the adaptive variance thresholding approach successfully enhanced overall model accuracy, the level of improvement varied among different AVT levels and ground-truth classes.

Furthermore, as evident from our experiments, the overall performance improvement on the external model was slightly superior to the internal model. The disparity may be attributed to the higher initial dimensionality of the external model, which allowed for less extensive pruning by adaptive variance thresholding. On the other hand, the internal model with lower initial dimensionality might have had fewer redundant features to begin with, hence a lesser margin for improvement. To gain further intuitive understanding, future work could explore the inverse process of identifying convolutional filter index based on the selected features post-AVT. This process could be an intriguing method to visualize the feature maps primarily associated with the Adaptive Variance Thresholding (AVT) sets.

Neural Architecture Search (NAS) played an instrumental role in this study, helping optimize the architecture of the classifier models for our specific task. The application of AVT resulted in a 60-fold reduction in the NAS input vector space at high levels. This improved the model's efficiency and expedited the inference speed and hyperparameter search. Despite these improvements, it is worth noting that we selected three threshold values placed at equal intervals. Although the prospect of an increased AVT sampling resolution could yield potentially superior results, it is essential to balance this with the considerably higher computational cost it would incur. In the Neural Architecture Search (NAS) process, we relied on validation accuracy as our primary evaluative metric. However, applying various evaluation metrics could potentially yield a diverse range of classifiers. Therefore, exploring potential meta-metrics which involve weighted combinations of various metrics may provide an edge in identifying architectures that are better tailored for specific tasks.

In our study, a significant constraint arose from the lack of label noise estimates for our dataset. This challenge is especially relevant for studies such as ours, focused on radiographic Knee Osteoarthritis (KOA). The early stages of KOA are usually shrouded in diagnostic uncertainty, thus making substantial label noise not just a mere possibility but a considerable expectation. Past research\cite{prezja2022deepfake} corroborated this, revealing significant discrepancies in labeling among medical experts. Without an estimate for label noise, our ability to discern the maximum achievable performance on the task is impeded. Moreover, another level of complexity stems from the potential for extensive feature noise, which may result from poor-quality radiographs. It remains unclear how our approach would respond under such circumstances, adding an extra dimension to the limitations faced in the study.

An additional limitation in our study pertained to the resolution parameters employed. Our evaluation was centered around transfer learning architectures and a resolution size of 224 x 224. This raises a question of scale adaptability: it is uncertain how our methodology would perform when applied to features from larger receptive fields in the feature learning segment. The influence of resolution size on model performance, especially when dealing with higher-resolution inputs, remains a topic unexplored within the confines of this study. This constitutes another key limitation that future studies would seek to address.

While our study was rooted in the context of KOA classification, our methodology's potential may extend beyond this specific domain. Given the inherent adaptability of AVT and NAS, our proposed paradigm has the potential to generalize to other areas of medicine or even entirely different fields that utilize deep transfer learning. However, such generalizations would necessitate extensive further research.

In conclusion, this study introduced a novel paradigm that enhanced the performance of existing deep-learning models for KOA classification. The proposed combination of AVT and NAS improved model accuracy and efficiency, and could potentially be applied to other models and diseases, presenting a promising avenue for future automatic diagnostic research.

\section*{Data Availability}
Trained models and the AVT code are available in the Google drive repository.

\bibliographystyle{elsarticle-num} 
\bibliography{main}

\section*{Acknowledgements}
The authors extend their sincere gratitude to Kimmo Riihiaho, Rodion Enkel and Leevi Lind for their exceptional support and invaluable discussions.

\section*{Author contributions statement}
Conceptualization: F. P.; 
Methodology: F. P.; 
Investigation: F. P.; 
Data curation: All authors; 
Formal analysis: All authors; 
Writing – original draft: F. P.; 
Writing – review \& editing: All authors.

\section*{Additional information}
 \textbf{Competing interests}
 All authors declare that they have no conflicts of interest.

\section*{Appendix}
\subsection*{EfficientNetV2M NAS}

\begin{table}[!ht]
\caption{NAS-selected architecture for the Low AVT Set of EfficientNet.}
\label{efilowavt}
\small
\begin{tabularx}{\textwidth}{lXX}
\toprule
\textbf{Layer Type} & \textbf{Output Shape} & \textbf{Parameter Count} \\
\midrule
Input Layer & 236 & - \\
Multi Category Encoding & 236 & - \\
Normalization & 236 & 473 \\
Dense Layer & 32 & 7584 \\
ReLU Activation & 32 & - \\
Dense Layer & 32 & 1056 \\
ReLU Activation & 32 & - \\
Dropout & 32 & - \\
Dense Layer & 5 & 165 \\
Softmax Activation & 5 & - \\
\bottomrule
\end{tabularx}
\end{table}

\begin{table}[!ht]
\caption{NAS-selected architecture for the Mid AVT Set of EfficientNet.}
\label{efimidavt}
\small
\begin{tabularx}{\textwidth}{lXX}
\toprule
\textbf{Layer Type} & \textbf{Output Shape} & \textbf{Parameter Count} \\
\midrule
Input Layer & 120 & - \\
Multi Category Encoding & 120 & - \\
Dense Layer & 16 & 1936 \\
ReLU Activation & 16 & - \\
Dense Layer & 32 & 544 \\
ReLU Activation & 32 & - \\
Dense Layer & 5 & 165 \\
Softmax Activation & 5 & - \\
\bottomrule
\end{tabularx}
\end{table}

\begin{table}[!ht]
\caption{NAS-selected architecture for the High AVT Set of EfficientNet.}
\label{efihighavt}
\small
\begin{tabularx}{\textwidth}{lXX}
\toprule
\textbf{Layer Type} & \textbf{Output Shape} & \textbf{Parameter Count} \\
\midrule
Input Layer & 4 & - \\
Multi Category Encoding & 4 & - \\
Normalization & 4 & 9 \\
Dense Layer & 32 & 160 \\
ReLU Activation & 32 & - \\
Dense Layer & 32 & 1056 \\
ReLU Activation & 32 & - \\
Dense Layer & 5 & 165 \\
Softmax Activation & 5 & - \\
\bottomrule
\end{tabularx}
\end{table}
\clearpage

\subsection{Chen VGG19 Ordinal NAS}
\begin{table}[!ht]
\caption{NAS-selected architecture for the Low AVT Set of the external Chen VGG19 Ordinal.}
\label{chenlowavtnas}
\small
\begin{tabularx}{\textwidth}{lXX}
\toprule
\textbf{Layer Type} & \textbf{Output Shape} & \textbf{Parameter Count} \\
\midrule
Input Layer & 4034 & - \\
Multi Category Encoding & 4034 & - \\
Normalization & 4034 & 8069 \\
Dense Layer & 32 & 129120 \\
Batch Normalization & 32 & 128 \\
ReLU Activation & 32 & - \\
Dense Layer & 32 & 1056 \\
Batch Normalization & 32 & 128 \\
ReLU Activation & 32 & - \\
Dense Layer & 5 & 165 \\
Softmax Activation & 5 & - \\
\bottomrule
\end{tabularx}
\end{table}

\begin{table}[!ht]
\caption{NAS-selected architecture for the Mid AVT Set of the external Chen VGG19 Ordinal.}
\label{chenmidavtnas}
\small
\begin{tabularx}{\textwidth}{lXX}
\toprule
\textbf{Layer Type} & \textbf{Output Shape} & \textbf{Parameter Count} \\
\midrule
Input Layer & 2048 & - \\
Multi Category Encoding & 2048 & - \\
Normalization & 2048 & 4097 \\
Dense Layer & 32 & 65568 \\
Batch Normalization & 32 & 128 \\
ReLU Activation & 32 & - \\
Dense Layer & 1024 & 33792 \\
Batch Normalization & 1024 & 4096 \\
ReLU Activation & 1024 & - \\
Dense Layer & 128 & 131200 \\
Batch Normalization & 128 & 512 \\
ReLU Activation & 128 & - \\
Dropout & 128 & - \\
Dense Layer & 5 & 645 \\
Softmax Activation & 5 & - \\
\bottomrule
\end{tabularx}
\end{table}

\begin{table}[!ht]
\caption{NAS-selected architecture for the High AVT Set of the external Chen VGG19 Ordinal.}
\label{chenhighavtnas}
\small
\begin{tabularx}{\textwidth}{lXX}
\toprule
\textbf{Layer Type} & \textbf{Output Shape} & \textbf{Parameter Count} \\
\midrule
Input Layer & 62 & - \\
Multi Category Encoding & 62 & - \\
Dense Layer & 128 & 8064 \\
ReLU Activation & 128 & - \\
Dropout & 128 & - \\
Dense Layer & 16 & 2064 \\
ReLU Activation & 16 & - \\
Dropout & 16 & - \\
Dense Layer & 32 & 544 \\
ReLU Activation & 32 & - \\
Dropout & 32 & - \\
Dense Layer & 5 & 165 \\
Softmax Activation & 5 & - \\
\bottomrule
\end{tabularx}
\end{table}





\end{document}